# Quasi phase matching in chiral materials

Bertrand Busson[1], Martti Kauranen[1*], Colin Nuckolls[2], Thomas J. Katz[2],
and André Persoons[1]

[1]*Laboratory of Chemical and Biological Dynamics and Center for Research on Molecular Electronics and Photonics, University of Leuven, B-3001 Heverlee, Belgium*

[2]*Department of Chemistry, Columbia University, New York, NY 10027, USA*

## Abstract

The second-order nonlinear optical coefficients associated with chirality differ in sign for the two mirror-image forms (enantiomers) of a chiral material. Structures comprised of alternating stacks of the enantiomers can therefore be used for quasi-phase-matched frequency conversion, as we demonstrate here by second-harmonic generation from Langmuir-Blodgett films of a helicenebisquinone. Such structures could lead to new types of frequency converters in which both the second-order nonlinear response and quasi phase matching arise from the chirality of a material rather than its polar order.

Frequency conversion of lasers is a significant application of second-order nonlinear optics. To do this efficiently, a material must fulfill two basic requirements. First, it must be noncentrosymmetric. Otherwise it will not have a second-order nonlinear response [1]. Second, it must allow the nonlinear source polarization and the generated field to propagate through it in phase. The interaction must be phase-matched [1].

In organic materials, molecular noncentrosymmetry is relatively easy to achieve. Macroscopic centrosymmetry, however, must usually be broken artificially. This can be done by various techniques, such as by aligning molecular dipoles in an external electric field (poling) or by depositing Langmuir-Blodgett or self-assembled films [2-4]. All these techniques result in polar order, the net alignment of molecular dipoles along an axis, the polar axis of the material. Such materials are often unstable, and their nonlinear response therefore decreases with time.

Phase matching is also hard to realize because the refractive indices of materials change with wavelength. In second-harmonic generation, for example, the wave vectors of the fundamental and second-harmonic beams $k_\omega$ and $k_{2\omega}$ are proportional to the refractive indices $n_\omega$ and $n_{2\omega}$ at the two frequencies. The phase (or wave vector) mismatch for copropagating fundamental and second-harmonic beams is thus $\Delta k = 2k_\omega - k_{2\omega} = 4\pi(n_\omega - n_{2\omega})/\lambda_\omega$, where $\lambda_\omega$ is the fundamental wavelength. Consequently, after a distance $L_c = \pi/|\Delta k|$, known as the coherence length, the source polarization and generated field are out of phase.

Perfect phase matching can be achieved in birefringent materials, because their refractive indices depend on the polarizations of the interacting fields [1]. However, in such materials the energy flux and wave propagation vectors usually are not parallel, and their usable thickness is thus limited. In addition, only rarely can their largest nonlinear coefficients be applied for phase matching.

Because of these limitations, an alternative to perfect phase matching, known as quasi phase matching, is frequently used [5,6]. In this procedure, the signs of the nonlinear optical coefficients are reversed after every coherence length (so-called lowest-order quasi phase matching), or after any odd multiple of it. This restores the phase relation between the source polarization and the generated field, allowing the nonlinear signal to grow quasi-continuously. Although the efficiency of quasi phase matching is lower than that of perfect phase matching, this is compensated by the ability to use the material's largest nonlinear coefficients. In polar materials, quasi phase matching is achieved by periodically inverting the direction of the polar axis, for example by poling the material in a periodic electric field [6].

However, polar order is not required for a material to be noncentrosymmetric [7,8]. Chiral molecules occur as two mirror-image forms (enantiomers) [9], and materials comprised of enantiomerically pure chiral molecules are always noncentrosymmetric. Accordingly, they give rise to second-order nonlinear optical effects even when there is no polar order [7,8], and the nonlinear coefficients associated with chirality can be large [8,10]. However, nonlinear processes relying on chirality have suffered from low overall efficiency because they have been observed only in experiments that are not phase-matched. But this limitation is not fundamental, because for enantiomers the nonlinear coefficients associated with chirality are opposite in sign [11]. This means that quasi-phase-matched frequency conversion should be possible in periodically alternating stacks of enantiomers.

In this Letter, we prepare such alternating stacks of enantiomers and demonstrate that they can be used for quasi-phase-matched second-harmonic generation. A series of samples was investigated in which equally thick stacks of the two enantiomers alternate. Quasi phase matching was verified by selecting the second-harmonic signal associated with chirality and measuring its intensity as a function of the thickness of the stacks. It was also shown that when the thickness of the stacks equals the coherence length, the second-harmonic signal grows continuously with the number of stacks.

To analyze the experiments, a model for the structure was considered that consists of $p$ stacks, each of thickness $L$, in which the sign of the second-order nonlinearity alternates (Fig. 1a). The equations of Fejer *et al.* [6] were used, but they were generalized to account for linear absorption at the second-harmonic wavelength. When the conversion efficiency is sufficiently low, the amplitude of the fundamental field can be assumed to be constant. The growth of the second-harmonic amplitude $E_2$ along the $z$ direction is then governed by the equation

$$\frac{dE_2}{dz} = -\alpha E_2 + \Gamma d(z) e^{i\Delta k z}, \tag{1}$$

where $\Gamma$ is a constant numerical factor proportional to the squared amplitude of the fundamental beam, $\alpha$ is the linear absorption coefficient at the second-harmonic wavelength, and $\Delta k$ is the wave vector mismatch. The nonlinear coefficient $d(z)$ reverses sign between successive stacks and is therefore given by $d(z) = (-1)^{j+1} d_0$ where $d_0$ is a constant and $j$, where $1 \leq j \leq p$, is the label of the stack.

Integrating Eq. (1) between $z = 0$ and $z = pL$ gives

$$E_2(pL) = \frac{\Gamma d_0}{\alpha + i\Delta k} \frac{1 - e^{-(\alpha + i\Delta k)L}}{1 + e^{-(\alpha + i\Delta k)L}} \left[ e^{-\alpha pL} - (-1)^p e^{i\Delta k pL} \right]. \tag{2}$$

In the particular case $L = L_c = \pi / |\Delta k|$, the intensity becomes

$$I_2(pL_c) = |E_2(pL_c)|^2 = \frac{|\Gamma d_0|^2}{\alpha^2 + (\Delta k)^2} \left( \frac{1 + e^{-\alpha L_c}}{1 - e^{-\alpha L_c}} \right)^2 (e^{-\alpha pL_c} - 1)^2. \tag{3}$$

According to Eq. (2) the output intensities from a series of films with fixed number of stacks $p$, but with increasing thickness $L$, are expected to grow until $L = L_c$ and then to decrease. When the thickness of each stack is equal to the coherence length, the intensity is expected to grow with the number of stacks, $p$, [Eq. (3)]. The growth should be quadratic when the material does not absorb the second-harmonic light ($\alpha \rightarrow 0$).

For our experiments, we used Langmuir-Blodgett films [12] of the helicenebisquinone shown in Fig. 1b [13]. Its pure enantiomers are known to aggregate in appropriate solvents and in Langmuir-Blodgett films into supramolecular helical arrays [13-15]. In Langmuir-Blodgett films, the nonlinear coefficients associated with chirality are large, as high as 50 pm/V, and approximately ten times larger than the coefficients not associated with chirality (that is, those that would not vanish even if the sample were achiral) [10].

The Langmuir-Blodgett films of the helicenebisquinone are known to be anisotropic in the plane of the substrate [10] and, when prepared by horizontal dipping, have $C_2$ symmetry, all molecular layers having the same orientation [10,14,16]. In this symmetry, both nonlinear coefficients that are associated with chirality and those that are not can be nonvanishing. However, the former account for most of the nonlinear response when the fundamental beam is *p*-polarized and the second-harmonic light detected is *s*-polarized. To further reduce the possible influence of the coefficients that are not associated with chirality, the samples were prepared by vertical dipping. This results in Y-type deposition [12], that is, deposition in which each molecular layer is oriented opposite to the previous one. Such films are expected to have $D_2$ symmetry, and therefore only coefficients associated with chirality nonvanishing [1]. Vertical dipping has the added benefit that it allows multilayers to be prepared more rapidly.

Films were prepared by spreading $2 \times 10^{-4}$ molar chloroform solutions of the enantiomers of the helicenebisquinone (the samples were more than 98% enantiomerically pure) onto a milli-Q water subphase (resistivity 18 MΩ cm) in a KSV trough. After the solvent had evaporated at room temperature, the monolayers were slowly compressed to a surface pressure of 20 mN m$^{-1}$. Films were deposited at low speed (10 mm/min) onto one side of hydrophobic glass substrates. The mean value of the transfer ratio (> 0.9, the same for the two enantiomers) indicates that the material deposits very well. Multilayers were prepared by depositing a stack with a given number of molecular layers of the (*P*)-(+)-enantiomer, following it by a stack with the same number of layers of the (*M*)-(-)-enantiomer, and repeating the sequence as necessary (Fig. 1a). Below, we refer to the samples by their sequence of stacks (e.g., *P*/*M*/*P*/*M*).

The Langmuir-Blodgett procedure allows films to be prepared with a thickness that can be controlled to a precision of one molecular layer. A drawback is that very large number of layers must be deposited to attain the thickness of even one coherence length, usually 1-10 μm for visible and near-infrared wavelengths. However, the coherence length is very much shorter if the second-harmonic light is detected in the reflected instead of the more common transmitted direction [17]. In this case, the wave vectors of the fundamental and second-harmonic fields are antiparallel, and the phase mismatch becomes $\Delta k = 2k_\omega + k_{2\omega} \sim 4k_\omega \sim 8\pi n_\omega / \lambda_\omega$. The coherence length is then *ca.* $L_c = \pi / \Delta k \sim \lambda_\omega / 8n_\omega \sim 100\,\text{nm}$. Thus the need to deposit very large numbers of molecular layers was avoided by conducting the experiments in the reflection geometry.

The experimental setup is shown in Fig. 2. A large (*ca.* 3 mm diameter) fundamental beam, provided by a Q-switched Nd:Yag laser ($\lambda_\omega$ = 1064 nm, 50 Hz repetition rate, 8 ns pulse length), irradiated the film sides of the samples. For reasons explained above, the fundamental beam was *p*-polarized with respect to the sample, and only *s*-polarized second-harmonic light (532 nm) was detected in the reflected direction. The experimental geometry is such that the signal grows along the surface normal [18], which is therefore the *z* axis in Eq. (1). This means that to analyze the results, the experimental parameters must be projected along this direction. The linear absorption coefficient of the samples at normal incidence was measured by a UV-visible spectrophotometer to be $\alpha = 1.21 \times 10^{-3}$ / molecular layer.

The first step was to determine the coherence length, $L_c$. This was done by measuring the second-harmonic response from a series of structures in which the thickness of the stacks was varied. In films comprised of only one enantiomer, such measurements could be hindered by interference between the part of the second-harmonic light emitted directly in the reflected direction and the part first emitted in the transmitted direction but then reflected by the film/substrate- and substrate/air interfaces. The latter part is absent in *P/M* samples because the coherence length in transmission is much larger than the thickness of the samples, and thus the opposite nonlinear responses of the *P* and *M* stacks cancel. In addition, the presence of equal amounts of the two enantiomers cancels any appreciable effects that could arise from the linear optical activity [9] of the material. Therefore, *P/M* samples were used for this measurement.

The experiments were carried out at two angles of incidence, 9.9° and 44.3°, and the data were fitted to Eq. (2) for *p* = *2* (Fig. 3). The agreement between the experimental data and the fitted curves is very good. The data thus show the expected characteristics of quasi phase matching. The coherence lengths obtained from the fitted curves for the two angles of incidence are 43.65 and 48.45 molecular layers, respectively. Assuming that the refractive indices at the fundamental and second-harmonic wavelengths are equal, these results imply that the refractive index of the helicenebisquinone films is 1.57 and that their molecular layers are 1.95 nm thick [19].

The experiments were repeated at the two angles of incidence previously used, but with a series of films comprised of four stacks (*P*/*M*/*P*/*M*). Since the absorption coefficient, refractive index, and coherence lengths for the two angles of incidence were now known, the data were fitted to Eq. (2) for $p = 4$ by using the amplitudes of the phase matching curves as the only adjustable parameter. The experimental and fitted results, also displayed in Fig. 3, are again in very good agreement.

Finally, when the thickness of each stack was one coherence length, the goal of quasi-phase-matched frequency conversion was reached: the second-harmonic intensity grew continuously with the number of stacks. Samples with 48 molecular layers per stack were used for the experiment. If the refractive index is taken to be 1.57, this thickness equals one coherence length when the angle of incidence is 42.6°. A new series of samples comprised of 1, 2, 3, 4 and 6 stacks was investigated at this angle of incidence. As displayed in Fig. 4, the experimental results follow Eq. (3), providing further evidence of quasi phase matching. The growth of the second-harmonic intensity with the number of stacks is somewhat less than quadratic only because the samples absorbed light at the second-harmonic wavelength.

In summary, the experiments show that quasi-phase-matched frequency conversion can be achieved in structures in which stacks of the two enantiomers of a chiral molecule alternate. In such structures, no polar order need be established, for both noncentrosymmetry and quasi phase matching are consequences of the chirality of the material. In the present work, quasi phase matching was demonstrated for second-harmonic generation in reflection. However, quasi phase matching by chirality should be limited neither to this particular process nor to the reflection geometry. Nor should it be limited to Langmuir-Blodgett films. It should apply to all frequency conversion processes, also in transmission, and to other structures in which enantiomers alternate. It should thereby make possible the design of completely new types of frequency converters.

We acknowledge the support of this work by the Belgian Government (IUAP P4/11), the Fund for Scientific Research - Flanders (FWOV G.0338.98, 9.0407.98), and the U.S. National Science Foundation (CHE9802316). We acknowledge contributions of S. Van Elshocht and T. Verbiest to early work that led to the experiments described. The AFM experiments were performed by G. de Schaetzen, the ellipsometric measurements by H. Bender. B. B. acknowledges the support of the European Union and M. K. that of the Academy of Finland.

* email: Martti.Kauranen@fys.kuleuven.ac.be, fax: +32 16 32 79 82

**Figure captions**

**Figure 1.**

(a) Structure of the films with alternating stacks of the two enantiomers (*P* and *M*). (b) Chemical structure of the helicenebisquinone.

**Figure 2.**

Experimental setup. The fundamental beam (1064 nm) is *p*-polarized with respect to the sample and applied at an angle of incidence $\theta$. The nonlinear interaction with the film produces second-harmonic light (532 nm). The *s*-polarized component of the second-harmonic signal is detected in the reflected direction using a photomultiplier.

**Figure 3.**

Experimental results for the second-harmonic intensities as functions of the total number of molecular layers of the *P/M* (circles) and *P/M/P/M* (squares) structures for (a) 9.9° and (b) 44.3° angles of incidence. Solid lines fit the experimental results to the theoretical function given by Eq. (2). The intensity scales of the four curves, arbitrary and independent of each other, are normalized to unity.

**Figure 4.**

Experimental results for the second-harmonic intensities as functions of the total number of molecular layers of the quasi-phase-matched structures. The thickness of each stack is 48 molecular layers, calculated to be equal to one coherence length at the 42.6° angle of incidence used in this measurement. Circles represent experimental points, the solid line fits them to the theoretical function given by Eq. (3).

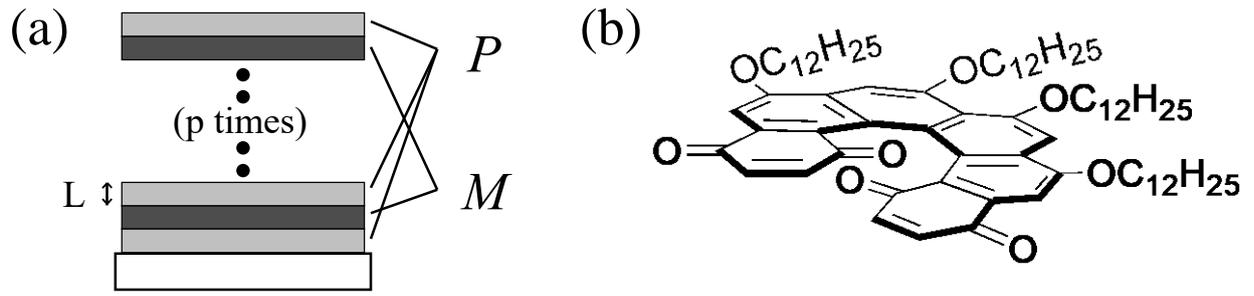

Figure 1

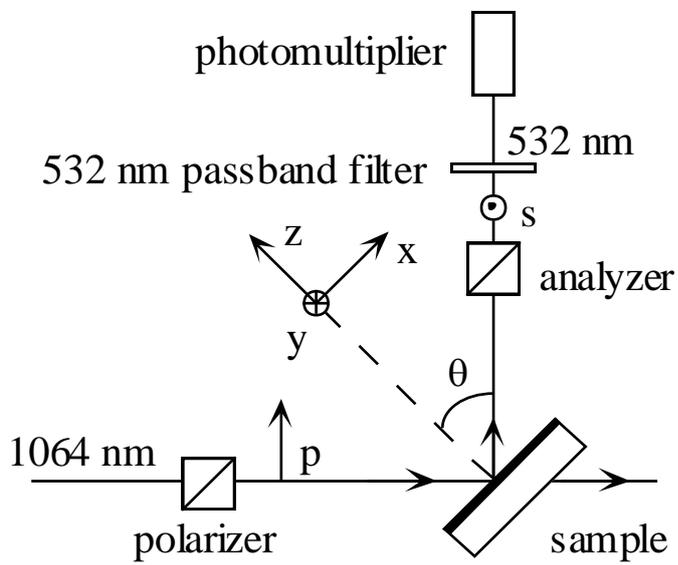

Figure 2

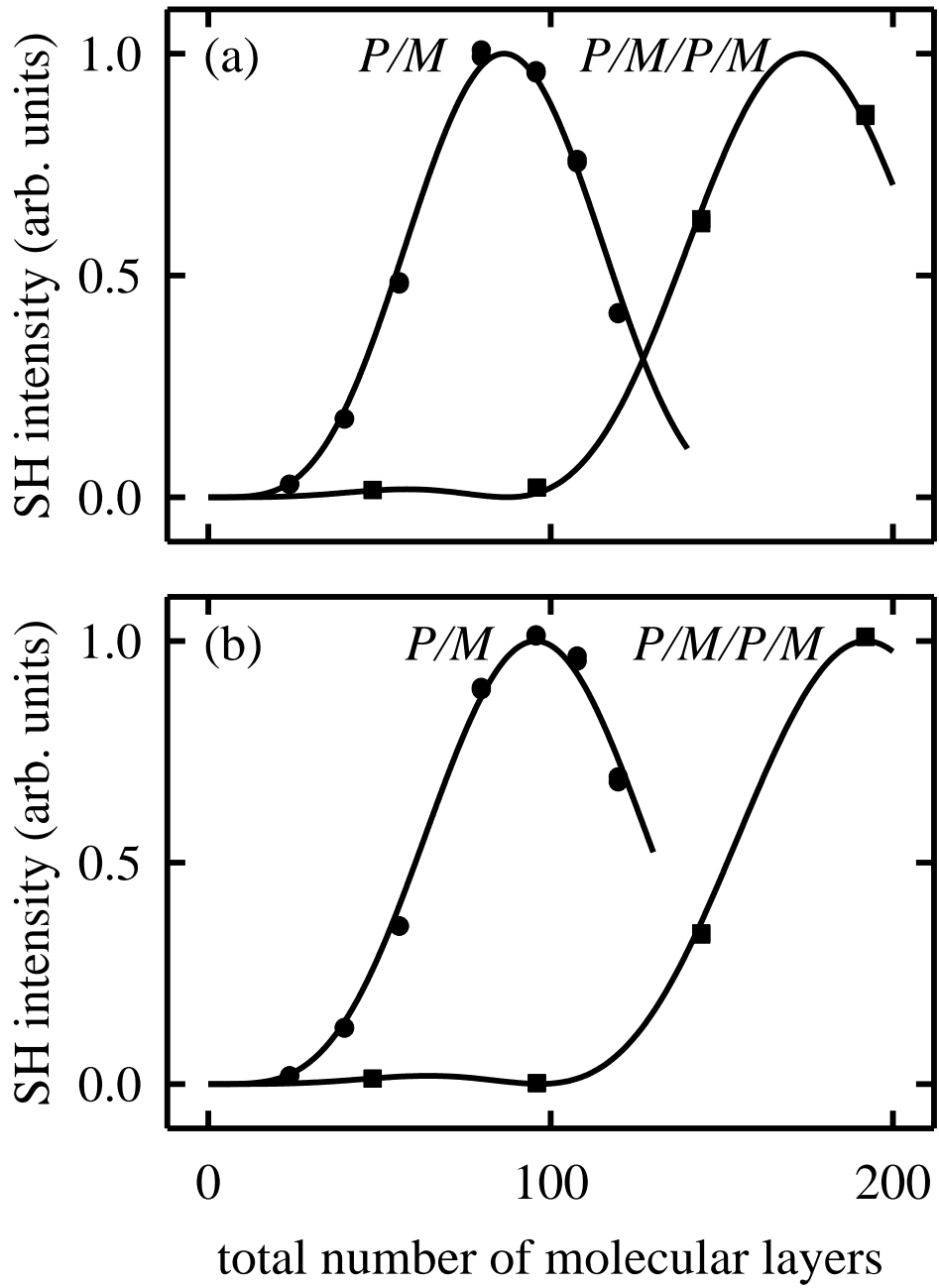

Figure 3

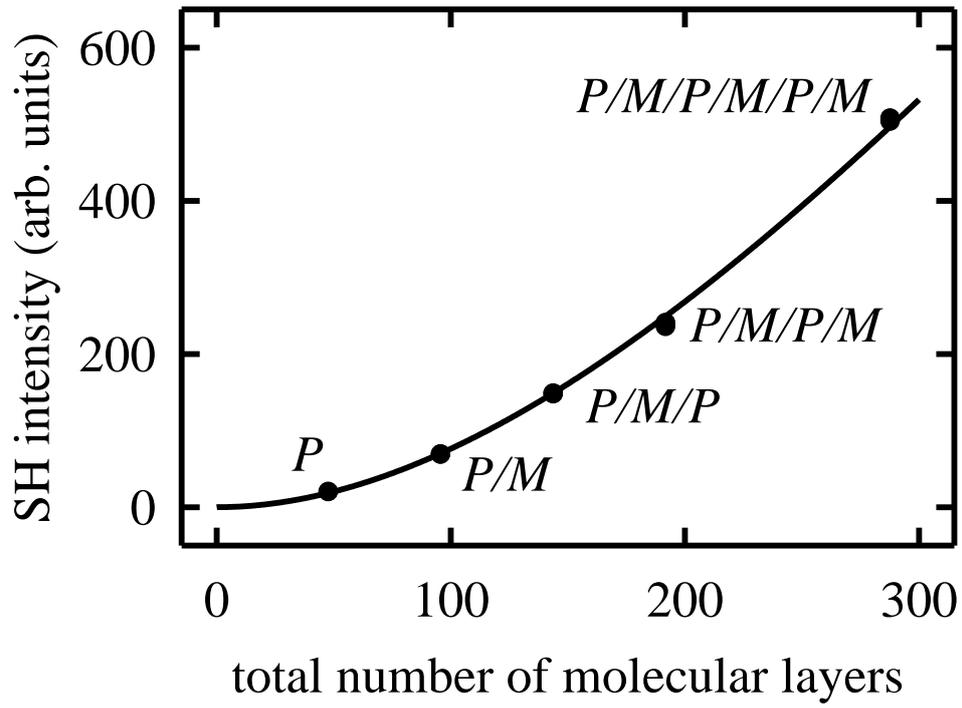

Figure 4